\documentclass[a4paper,11pt]{article}
\usepackage{pos}
\usepackage{amsmath}
\usepackage{tikz}
\usetikzlibrary{positioning}
\usetikzlibrary{patterns}
\newcommand{\dd}{\textrm{d}}

\title{Taming a resurgent ultra-violet renormalon}

\author[a,1]{Michael Borinsky}
\author*[b]{David Broadhurst}

\affiliation[a]{Institute for Theoretical Studies, ETH Z\"urich\\
8092 Z\"urich, Switzerland}

\affiliation[b]{School of Physical Sciences, Open University\\
Milton Keynes MK7 6AA, UK}

\note{ Supported by Dr.\ Max R\"ossler, the Walter Haefner Foundation and the ETH Z\"urich Foundation. }

\emailAdd{Michael.Borinsky@eth-its.ethz.ch}
\emailAdd{David.Broadhurst@open.ac.uk}

\abstract{Perturbative expansions in quantum field  theory diverge for at least two reasons:
the number of Feynman diagrams increases dramatically with the loop number
and the process of renormalization may make the contribution of some diagrams large.
We give an example of the second problem, from an ultra-violent renormalon of
$\phi^3$ theory in 6 dimensions, where we can compute to very high loop-order.
Taming this renormalon involves recent work on resurgence. 
This challenge is much more demanding than the corresponding problem for Yukawa
theory in 4 dimensions.}

\FullConference{
  Loops and Legs in Quantum Field Theory - LL2022,\\
  25-30 April, 2022\\
  Ettal, Germany
}

\begin{document}
\maketitle

\section{Dyson-Schwinger equation and asymptotic expansion}

Consider the perturbation expansion generated by the iteration of a single divergent diagram, via the non-linear Dyson-Schwinger equation
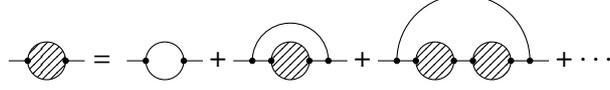
\cite{BK1,BK2,BDM,BB},
\begin{align}
\begin{tikzpicture}[baseline={([yshift=1ex]current bounding box.south)}] \coordinate (in); \coordinate[right=.25 of in] (v1); \coordinate[right=.25 of v1] (m); \coordinate[right=.25 of m] (v2); \coordinate[right=.25 of v2] (out); \filldraw[preaction={fill,white},pattern=north east lines] (m) circle(.25); \filldraw (v1) circle(1pt); \filldraw (v2) circle(1pt); \draw (in) -- (v1); \draw (v2) -- (out); \end{tikzpicture}
=
~
\begin{tikzpicture}[baseline={([yshift=1ex]current bounding box.south)}] \coordinate (in); \coordinate[right=.25 of in] (v1); \coordinate[right=.25 of v1] (vm); \coordinate[right=.25 of vm] (v2); \coordinate[right=.25 of v2] (out); \filldraw (v1) circle(1pt); \filldraw (v2) circle(1pt); \draw (in) -- (v1); \draw (v2) -- (out); \draw (vm) circle (.25); \end{tikzpicture}
+
\begin{tikzpicture}[baseline={([yshift=1ex]current bounding box.south)}] \coordinate (in); \coordinate[right=.25 of in] (v1); \coordinate[right=.25 of v1] (v2); \coordinate[right=.25 of v2] (m); \coordinate[right=.25 of m] (v3); \coordinate[right=.25 of v3] (v4); \coordinate[right=.25 of v4] (out); \filldraw[preaction={fill,white},pattern=north east lines] (m) circle(.25); \filldraw (v1) circle(1pt); \filldraw (v2) circle(1pt); \filldraw (v3) circle(1pt); \filldraw (v4) circle(1pt); \draw (in) -- (v1); \draw (v1) -- (v2); \draw (v3) -- (v4); \draw (v4) -- (out); \draw (v1) arc (180:0:.5); \end{tikzpicture}%
+
\begin{tikzpicture}[baseline={([yshift=1ex]current bounding box.south)}] \coordinate (in); \coordinate[right=.25 of in] (v1); \coordinate[right=.25 of v1] (v2); \coordinate[right=.25 of v2] (m1); \coordinate[right=.25 of m1] (v3); \coordinate[right=.25 of v3] (v4); \coordinate[right=.25 of v4] (m2); \coordinate[right=.25 of m2] (v5); \coordinate[right=.25 of v5] (v6); \coordinate[right=.25 of v6] (out); \filldraw[preaction={fill,white},pattern=north east lines] (m1) circle(.25); \filldraw[preaction={fill,white},pattern=north east lines] (m2) circle(.25); \filldraw (v1) circle(1pt); \filldraw (v2) circle(1pt); \filldraw (v3) circle(1pt); \filldraw (v4) circle(1pt); \filldraw (v5) circle(1pt); \filldraw (v6) circle(1pt); \draw (in) -- (v1); \draw (v1) -- (v2); \draw (v3) -- (v4); \draw (v5) -- (v6); \draw (v6) -- (out); \draw (v1) arc (180:0:.875); \end{tikzpicture}%
+
\cdots\nonumber
\end{align} 
contributing to the self-energy term $\Sigma$ in the  inverse propagator $q^2(1-\Sigma)$, for a massless scalar 
particle with a $\phi^3$ interaction, in the critical space-time dimension $D=6$,  for which the  coupling constant is dimensionless. 

The dependence of $\Sigma$ on the external momentum $q$ comes solely from renormalization.  At $n$ loops,
we get a contribution that is a polynomial of degree $n$ in $\log(q^2/\mu^2)$, 
multiplied by $a^n$ where
$a=\lambda^2/(4\pi)^3$, $\lambda$ is the coupling constant and $\mu$ is the renormalization scale.

If we use momentum-space subtraction, so that $\Sigma$ vanishes at $q^2=\mu^2$, the dependence
on momentum is completely determined by the anomalous dimension, with
\begin{equation}\gamma(a) = -\left.q^2\frac{{\rm d}\Sigma}{{\rm d}q^2}\right|_{q^2=\mu^2}\quad\text{giving}\quad
\left.\frac{{\rm d}\log(1-\Sigma)}{{\rm d}\log q^2}\right|_{q^2=\mu^2}=\gamma\left(\frac{a}{(1-\Sigma)^2}\right).
\end{equation} 
 
The number of distinct diagrams at $n$ loops is the number $T_n$ of rooted trees with $n$ nodes,
which gives the sequence~\cite{polya1937kombinatorische},
$$1,\, 1,\, 2,\, 4,\, 9,\, 20,\, 48,\, 115,\, 286,\, 719,\, 1842,\, 4766,\, 12486,\, 32973,\, 87811,\, 235381,\ldots$$
up to 16 loops.
The iterated structure: tree = root + branches, with every branch being itself a tree,
gives the asymptotic growth
\begin{gather}
T_{n}=\frac{b}{n^{3/2}}c^n(1+O(1/n))\\
b = 0.43992401257102530404090339143454476479808540794011\ldots\\
c = 2.95576528565199497471481752412319458837549230466359 \ldots
\end{gather}
At 250 loops, the number of Feynman diagrams is
\begin{verbatim}
T_250=517763755754613310897899496398412372256908589980657316
271041790137801884375338813698141647334732891545098109934676.
\end{verbatim}
This is not the main source of the problem.  If the contribution of each
diagram was bounded,
there would be a finite radius of convergence for the perturbation expansion.
The divergence of the series comes from renormalization, which makes
the  $n$-loop term grow factorially. This is called a {\em renormalon} singularity~\cite{Lautrup:1977hs}.

At 4 loops, we have a rainbow, a chain and two more interesting diagrams:
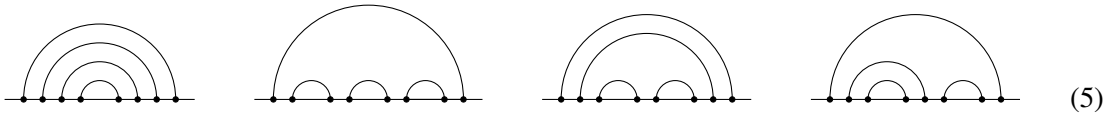
\begin{align}
\label{diagrams4}
\begin{tikzpicture}[baseline={([yshift=-.6ex]current bounding box.center)}] \coordinate (in); \coordinate[right=.25 of in] (v1); \coordinate[right=.25 of v1] (v2); \coordinate[right=.25 of v2] (v3); \coordinate[right=.25 of v3] (v4); \coordinate[right=.25 of v4] (m); \coordinate[right=.25 of m] (v5); \coordinate[right=.25 of v5] (v6); \coordinate[right=.25 of v6] (v7); \coordinate[right=.25 of v7] (v8); \coordinate[right=.25 of v8] (out); \draw[white] (v1) arc (-180:0:1); \filldraw (v1) circle(1pt); \filldraw (v2) circle(1pt); \filldraw (v3) circle(1pt); \filldraw (v4) circle(1pt); \filldraw (v5) circle(1pt); \filldraw (v6) circle(1pt); \filldraw (v7) circle(1pt); \filldraw (v8) circle(1pt); \draw (in) -- (v1); \draw (v1) -- (v2); \draw (v2) -- (v3); \draw (v3) -- (v4); \draw (v4) -- (v5); \draw (v5) -- (v6); \draw (v6) -- (v7); \draw (v7) -- (v8); \draw (v8) -- (out); \draw (v1) arc (180:0:1); \draw (v2) arc (180:0:.75); \draw (v3) arc (180:0:.5); \draw (v4) arc (180:0:.25); \end{tikzpicture}
\qquad
\begin{tikzpicture}[baseline={([yshift=-.6ex]current bounding box.center)}] \coordinate (in); \coordinate[right=.25 of in] (v1); \coordinate[right=.25 of v1] (v2); \coordinate[right=.25 of v2] (m1); \coordinate[right=.25 of m1] (v3); \coordinate[right=.25 of v3] (v4); \coordinate[right=.25 of v4] (m2); \coordinate[right=.25 of m2] (v5); \coordinate[right=.25 of v5] (v6); \coordinate[right=.25 of v6] (m3); \coordinate[right=.25 of m3] (v7); \coordinate[right=.25 of v7] (v8); \coordinate[right=.25 of v8] (out); \draw[white] (v1) arc (-180:0:1.25); \filldraw (v1) circle(1pt); \filldraw (v2) circle(1pt); \filldraw (v3) circle(1pt); \filldraw (v4) circle(1pt); \filldraw (v5) circle(1pt); \filldraw (v6) circle(1pt); \filldraw (v7) circle(1pt); \filldraw (v8) circle(1pt); \draw (in) -- (v1); \draw (v1) -- (v2); \draw (v2) -- (v3); \draw (v3) -- (v4); \draw (v4) -- (v5); \draw (v5) -- (v6); \draw (v6) -- (v7); \draw (v7) -- (v8); \draw (v6) -- (out); \draw (v1) arc (180:0:1.25); \draw (v2) arc (180:0:.25); \draw (v4) arc (180:0:.25); \draw (v6) arc (180:0:.25); \end{tikzpicture}
\qquad
\begin{tikzpicture}[baseline={([yshift=-.6ex]current bounding box.center)}] \coordinate (in); \coordinate[right=.25 of in] (v0); \coordinate[right=.25 of v0] (v1); \coordinate[right=.25 of v1] (v2); \coordinate[right=.25 of v2] (m1); \coordinate[right=.25 of m1] (v3); \coordinate[right=.25 of v3] (v4); \coordinate[right=.25 of v4] (m2); \coordinate[right=.25 of m2] (v5); \coordinate[right=.25 of v5] (v6); \coordinate[right=.25 of v6] (v7); \coordinate[right=.25 of v7] (out); \draw[white] (v1) arc (-180:0:1.125); \filldraw (v0) circle(1pt); \filldraw (v1) circle(1pt); \filldraw (v2) circle(1pt); \filldraw (v3) circle(1pt); \filldraw (v4) circle(1pt); \filldraw (v5) circle(1pt); \filldraw (v6) circle(1pt); \filldraw (v7) circle(1pt); \draw (in) -- (v1); \draw (v1) -- (v2); \draw (v2) -- (v3); \draw (v3) -- (v4); \draw (v4) -- (v5); \draw (v5) -- (v6); \draw (v6) -- (out); \draw (v0) arc (180:0:1.125); \draw (v1) arc (180:0:.875); \draw (v2) arc (180:0:.25); \draw (v4) arc (180:0:.25); \end{tikzpicture}%
\qquad
\begin{tikzpicture}[baseline={([yshift=-.6ex]current bounding box.center)}] \coordinate (in); \coordinate[right=.25 of in] (v1); \coordinate[right=.25 of v1] (v2); \coordinate[right=.25 of v2] (v3); \coordinate[right=.25 of v3] (m1); \coordinate[right=.25 of m1] (v4); \coordinate[right=.25 of v4] (v5); \coordinate[right=.25 of v5] (v6); \coordinate[right=.25 of v6] (m2); \coordinate[right=.25 of m2] (v7); \coordinate[right=.25 of v7] (v8); \coordinate[right=.25 of v8] (out); \draw[white] (v1) arc (-180:0:1.125); \filldraw (v1) circle(1pt); \filldraw (v2) circle(1pt); \filldraw (v3) circle(1pt); \filldraw (v4) circle(1pt); \filldraw (v5) circle(1pt); \filldraw (v6) circle(1pt); \filldraw (v7) circle(1pt); \filldraw (v8) circle(1pt); \draw (in) -- (v1); \draw (v1) -- (v2); \draw (v2) -- (v3); \draw (v3) -- (v4); \draw (v4) -- (v5); \draw (v5) -- (v6); \draw (v6) -- (out); \draw (v1) arc (180:0:1.125); \draw (v2) arc (180:0:.5); \draw (v3) arc (180:0:.25); \draw (v6) arc (180:0:.25); \end{tikzpicture}%
\end{align}

The sum of rainbows converges. Chains can be summed  by Borel transformation~\cite{BK1}.
\begin{eqnarray}\gamma_{\rm rainbow}=\frac{3-\sqrt{5+4\sqrt{1+a}}}{2}
&=&-\frac{a}{6}+11\,\frac{a^2}{6^3}-206\,\frac{a^3}{6^5}+4711\,\frac{a^4}{6^7}+O(a^5)\\
\gamma_{\rm chain}=-\int_0^\infty\frac{6\exp(-6z/a)\dd z}{(z+1)(z+2)(z+3)}
&=&-\frac{a}{6}+11\,\frac{a^2}{6^3}-170\,\frac{a^3}{6^5}+3450\,\frac{a^4}{6^7}+O(a^5)\\
\gamma=\sum_{n>0}G_n\,\frac{(-a)^n}{6^{2n-1}}
&=&-\frac{a}{6}+11\,\frac{a^2}{6^3}-376\,\frac{a^3}{6^5}+20241\,\frac{a^4}{6^7}+O(a^5)
\end{eqnarray}
with large integers $G_n$ in the {\em alternating} asymptotic series for $\gamma$. 
Note that $G_4=20241>4711+3450$, because of two further diagrams in \eqref{diagrams4}.
In one we have a chain inside a double rainbow. In the other, a double rainbow is chained
with the primitive divergence. This interplay is coded by rooted trees.

At 500 loops, the integer coefficient $G_{500}$
has 1675 decimal digits. It was determined in work of the second author with Dirk Kreimer that resulted in a third-order differential equation \cite{BK2},
\begin{gather}
8a^3\gamma\left\{\gamma^2\gamma^{\prime\prime\prime}
+4\gamma\gamma^\prime\gamma^{\prime\prime}
+(\gamma^\prime)^3\right\}
+4a^2\gamma\left\{2\gamma(\gamma-3)\gamma^{\prime\prime}
+(\gamma-6)(\gamma^\prime)^2\right\}
\nonumber\\{}
+2a\gamma(2\gamma^2+6\gamma+11)\gamma^\prime
-\gamma(\gamma+1)(\gamma+2)(\gamma+3)=a
\end{gather}
with quartic non-linearity. 

Interest in this problem came from Kreimer's discovery of the Hopf algebra of the
iterated subtraction of subdivergences~\cite{Connes:1999yr}, whose utility was illustrated in this
example, with a single primitive divergence leading to undecorated rooted trees.

The corresponding diagrams in Yukawa theory, in its critical dimension $D=4$, give
a  first-order equation with merely quadratic non-linearity, which was solved using
the complementary error function \cite{BK2}, thereby achieving explicit all-orders
results for both the anomalous dimension and the self-energy. The expansion coefficients
in this simpler case enumerate  connected chord diagrams and an all-order resurgence analysis is possible~\cite{BD}.

The $D=4$ and $D=6$ examples were also investigated in the more cumbersome
minimal subtraction scheme, where one retains finite parts of 
$\Sigma$ at $q^2=\mu^2$. Here one encounters unwieldy products of zeta values
with weights that increase linearly with the loop-number.
Recently,  Paul-Hermann Balduf has shown how to absorb these into
a rescaling of $\mu$ that can be expanded in the coupling $a$~\cite{Balduf:2021kag}. 

\section{Pad\'e-Borel summation with alternating signs}

Broadhurst and Kreimer resummed the factorially divergent alternating series by an Ansatz~\cite{BK1}
\begin{equation}\gamma(a)= -\frac{a}{6\Gamma(\beta)}\int_0^\infty P(ax/3)\exp(-x)x^{\beta-1}{\rm d}x,
\quad P(z)=\frac{N(z)}{D(z)}.
\end{equation}
The expansion coefficients of $P(z)=1+O(z)$ are obtained from those
those of $\gamma(a)/a$ by dividing the latter by factorially increasing factors, producing
a function $P$ which was expected to have a finite radius of convergence in the Borel variable $z$,
with singularities on the negative $z$-axis, as for the sum of chains.

The Pad\'e trick is to convert the expansion of $P$, up to $n$ loops, into a ratio $N/D$ of polynomials
of degrees close to $n/2$. Then one can check how well this method reproduces $G_{n+1}$.
It was found that this works rather well with $\beta\approx3$ .
For example, fitting the first 29 values of $G_n$ with a ratio of polynomials of degree 14
gave a pole, coming from the denominator $D(z)$, at $z=-0.994$. The other 13 poles occurred 
further to the left, with $\Re z<-1$. Moreover the numerator $N(z)$ gave no zero with $\Re z > 0$.
Then this method reproduced the first 15 decimal digits of $G_{30}$. 

The first author, Gerald Dunne and Max Meynig have
recently shown that this method works even better with $\beta=\frac{35}{12}$~\cite{BDM}, for reasons
that we now explain.

\section{Trans-series and resurgent hyperasymptotics}

There is an old and rather loose argument, going back to Freeman Dyson in 1952~\cite{Dyson:1952tj}, that we should {\em not} expect
realistic field theories to give convergent expansions in the square of a coupling constant.
If they did, we could get sensible answers for a non-unitary theory with an imaginary coupling constant,
such as an electrodynamics in which electrons repel positrons.

There is an amusing converse of this suggestion. If you find an expansion
that is Borel summable,  then study it at imaginary coupling. In the case of $\phi^3$ theory the resulting non-unitary theory relates
to the Yang-Lee edge singularity in condensed matter physics~\cite{Fisher_1978}.

So now we recast the Broadhurst-Kreimer problem, in the manner of Borinksy, Dunne and Meynig~\cite{BDM},
by setting $g(x)=\gamma(-3x)/x$, to obtain an ODE that is economically written as 
\begin{equation}
\label{ode}
(g(x)P-1)(g(x)P-2)(g(x)P-3)g(x)=-3,\quad  P=x\left(2x\frac{\dd}{{\dd}x}+1\right),
\end{equation}
and has an  unsummable formal perturbative solution 
\begin{equation}
g_0(x)\sim\sum_{n=0}^\infty A_nx^n=\frac{1}{2}+\frac{11}{24}x
+\frac{47}{36}x^2+\frac{2249}{384}x^3+\frac{356789}{10368}x^4
+\frac{60819625}{248832}x^5+O(x^6).
\end{equation}
The expansion coefficients behave as
\begin{equation}
A_n=S_1\Gamma\left(n+\frac{35}{12}\right)\left(1-\frac{97}{48}\left(\frac{1}{n}\right)+O\left(\frac{1}{n^2}\right)\right),
\end{equation}
at large $n$, with a Stokes constant
\begin{equation}
S_1=0.087595552909179124483795447421262990627388017406822\ldots
\end{equation}
that can be determined, empirically, by considering a solution
\begin{equation}
g(x)=g_0(x)+\sigma_1 x^{-\beta}\exp(-1/x) h_1(x)+O(\sigma_1^2)
\end{equation}
and retaining terms linear in $\sigma_1$ in the non-linear ODE. This yields a linear homogeneous ODE for $h_1(x)$,
which permits a solution that is  finite and regular at $x=0$ if and only if $\beta=\frac{35}{12}$. 
Normalizing $\sigma_1$ by setting $h_1(0)=-1$, we obtain the expansion of
\begin{equation}h_1(x)\sim\sum_{k=0}^\infty B_kx^k=
-1 + \frac{97}{48}x + \frac{53917}{13824}x^2+\frac{3026443}{221184}x^3 + \frac{32035763261}{382205952}x^4
+O(x^5)
\end{equation}
which gives the  first-instanton correction to the perturbative solution, suppressed by $\exp(-1/x)$.
 
By developing the series $A_n$ and $B_k$, we were able to determine 3000 digits of $S_1$~\cite{BB} in
\begin{equation}A_n\sim- S_1\sum_{k\ge0}\Gamma\left(n+\frac{35}{12}-k\right)B_k.\end{equation} 
This is an example of {\em resurgence}~\cite{ecalle1981fonctions}: information about $A_n$ resurges in $B_k$, and vice versa,
because both $A(x)=g_0(x)$ and $B(x)=h_1(x)$ know about the same physics.

Hyperasymptotic expansions concern the study of how $B_n$ behaves at large $n$, which involves another
set of numbers $C_k$, at small $k$, and so on, {\em ad infinitum}. They
involve terms suppressed by $\exp(-m/x)$, with action $m>1$. For this third-order ODE,
there are 3 solutions to the linearized problem~\cite{BDM}, namely
\begin{equation}
g(x)=g_0(x)+\sigma_m\left(x^{-\frac{35}{12}}e^{-\frac{1}{x}}\right)^m h_m(x)+O(\sigma_m^2),\quad m\in \{1,2,3\},
\end{equation}
with $h_2/x^5=C$ and $h_3/x^5=D$ finite and regular near the origin. 

Then we use a linearized ODE to develop the expansions
\begin{align}
C(x)&=h_2(x)/x^5=
-1 + \frac{151}{24}x - \frac{63727}{3456}x^2 + \frac{7112963}{82944}x^3 - \frac{7975908763x}{23887872}x^4
+O(x^5),\\
D(x)&=h_3(x)/x^5=
-1 + \frac{227}{48}x + \frac{1399}{4608}x^2 + \frac{814211}{73728}x^3 + \frac{3444654437}{42467328}x^4
+O(x^5).
\end{align}

This suggests to also study the higher order corrections in $\sigma_1,\sigma_2$ and $\sigma_3$, which leads to the \emph{trans-series}. The trans-series organizes the higher order instanton corrections to the solution of the ODE and it neatly reflects the perpetuating low-order/large-order correspondence of the hyperasymptotic expansions.
Before presenting the trans-series solution to~\eqref{ode}, we remark on some of its general features.
\begin{enumerate}
\item
The terms suppressed by $\exp(-2/x)$ involve $\sigma_2$ and $\sigma_1^2$. The former are given by $C$ and the latter
are determined by an inhomogeneous linear ODE, whose solution is ambiguous, up to a multiple of 
the homogeneous solution $h_2=x^5C$,  since we can shift $\sigma_2$ by a multiple of $\sigma_1^2$. 
\item
In the terms suppressed by $\exp(-3/x)$ there a second ambiguity, since we can shift $\sigma_3$ by a multiple of $\sigma_1^3$. 
\item
Ambiguities of inhomogeneous solutions occur at places in expansions where logarithms first arise.
This happens when the power of $x$ in an expansion is a multiple of 5.
\item The highest power of
$\log(x)$, in terms with action $m$, is $\lfloor m/2\rfloor$.
\end{enumerate}

The terms in the  trans-series solution to~\eqref{ode} with action $m\le4$ are of the form~\cite{BB},
\begin{gather}
g=\sum_{m\ge0}g_m\left(x^{-\frac{35}{12}}\,e^{-\frac{1}{x}}\right)^m,\quad L=\frac{21265}{2304}x^5\log(x),\\
g_0=A,\quad g_1=\sigma_1B,\quad g_2=\sigma_2x^5C+\sigma_1^2(F+CL),\\
g_3=\sigma_3x^5D+\sigma_1\sigma_2x^5E+\sigma_1^3(I+(D+E)L),\\
g_4=\sigma_1\sigma_3x^5G+\sigma_2^2x^{10}H+\sigma_1^2\sigma_2x^5(J+2HL)
+\sigma_1^4(K+(G+J)L+HL^2).
\end{gather}
Denoting the coefficients of $x^n$ in functions by subscripts, we found that the choices
\begin{equation}\frac{F_5}{2!}=\frac{I_5}{3!}=\frac{32642693907919}{36691771392}\end{equation}
greatly simplify our system of hyperasymptotic expansions. Then
\begin{gather}
B_n\sim-2S_1\sum_{k\ge0}F_k\Gamma(n+\tfrac{35}{12}-k)\nonumber\\{}
+4S_1\sum_{k\ge0}C_k\Gamma(n-\tfrac{25}{12}-k)\left(\tfrac{21265}{4608}\psi(n-\tfrac{25}{12}-k)+d_1\right),\\
d_1=-43.332634728250755924500717390319380703460728022278\ldots
\end{gather}
with $\psi(z)=\Gamma{^\prime}(z)/\Gamma(z)=\log(z)+O(1/z)$, shows the $m=1$ term, at large $n$,
looking forward to $m=2$ terms, at small $k$.

For the asymptotic expansion of the second-instanton coefficients, we found
\begin{equation}
C_n\sim-S_1\sum_{k\ge0}E_k\Gamma(n+\tfrac{35}{12}-k)+S_3\sum_{k\ge0}B_k(-1)^{n-k}\Gamma(n+\tfrac{25}{12}-k).\label{Cn}
\end{equation}
The first sum looks forwards to $m=3$ in the trans-series, where coefficients of
\begin{equation}
E(x)=-4 + \tfrac{371}{12}x - \tfrac{111785}{1152}x^2 + \tfrac{8206067}{18432}x^3 - \tfrac{18251431003}{10616832}x^4
+O(x^5)
\end{equation}
appear. It does not contain the coefficients $D_k$ of the third instanton, which decouples
from the asymptotic expansion for the second instanton.

The second sum in~(\ref{Cn}) has alternating signs, looks backwards to $m=1$ and is
suppressed by a factor of $1/n^{5/6}$.  This can be understood using alien calculus. Likewise,
\begin{align}
F_n\sim&-3S_1\sum_{k\ge0}I_k\Gamma(n+\tfrac{35}{12}-k)\nonumber\\
&+2S_1\sum_{k\ge0}(3D_k+2E_k)\Gamma(n-\tfrac{25}{12}-k)\left(\tfrac{21265}{4608}\psi(n-\tfrac{25}{12}-k)+d_1\right)\nonumber\\
&-2S_3\sum_{k\ge0}B_k(-1)^{n-k}\Gamma(n-\tfrac{35}{12}-k)\left(\tfrac{21265}{4608}\psi(n-\tfrac{35}{12}-k)+f_1\right)\label{Fn}
\end{align} 
looks forwards to $I_k$, $D_k$ and $E_k$, at $m=3$, and backwards to $B_k$ at, $m=1$.

The new constants in~(\ref{Fn}) are
\begin{eqnarray}
S_3&=&2.1717853140590990211608601227903892302479464193027\ldots\\
f_1&=&-40.903692509228515003814479126901354785263669553014\ldots
\end{eqnarray}
Two more were discovered in the  backward looking terms of
\begin{gather}
I_n\sim-4S_1\sum_{k\ge0}K_k\Gamma(n+\tfrac{35}{12}-k)\nonumber\\
+2S_1\sum_{k\ge0}(3G_k+2J_k)\Gamma(n-\tfrac{25}{12}-k)\left(\tfrac{21265}{4608}\psi(n-\tfrac{25}{12}-k)+d_1\right)\nonumber\\
-4S_3\sum_{k\ge0}F_k(-1)^{n-k}\Gamma(n-\tfrac{35}{12}-k)\left(\tfrac{21265}{4608}\psi(n-\tfrac{35}{12}-k)+f_1\right)\nonumber\\
-8S_3\sum_{k\ge0}C_k(-1)^{n-k}\Gamma(n-\tfrac{95}{12}-k)Q(n-\tfrac{95}{12}-k),\\
Q(z)=\left(\tfrac{21265}{4608}\right)^2\left(\psi^2(z)+\psi^\prime(z)\right)
+2c_1\left(\tfrac{21265}{4608}\right)\psi(z)+c_2,\\
c_1=-41.031956764302710583921068101545509453704897898188\ldots\\
c_2/c_1^2=1.0002016472131992595822805380838324188011572304276\ldots
\end{gather}
We believe that 6 constants, $S_1,d_1,S_3,f_1,c_1,c_2$, suffice for the complete description of resurgence.

{\bf Conjecture}~\cite{BB}: The trans-series solution to~\eqref{ode} and its resurgence take the forms
\begin{gather}
g(x)=\sum_{m=0}^\infty\left(x^{-\frac{35}{12}}\,e^{-\frac{1}{x}}\right)^m
\sum_{i=0}^{\lfloor m/2\rfloor}\sum_{j=0}^{\lfloor (m-2i)/3\rfloor}
\sigma_1^{m-2i-3j}\widehat{\sigma}_2^i\widehat{\sigma}_3^jx^{5(i+j)}\sum_{n\ge0}a^{(m)}_{i,j}(n)x^n,\\
\widehat{\sigma}_2=\sigma_2+\tfrac{21265}{2304}\sigma_1^2\log(x),\quad
\widehat{\sigma}_3=\sigma_3+\tfrac{21265}{2304}\sigma_1^3\log(x),\\[2pt]
a^{(m)}_{i,j}(n)\sim-(s+1)S_1\sum_{k\ge0}a^{(m+1)}_{i,j}(k)\Gamma(n+\tfrac{35}{12}-k)\nonumber\\
{}+S_1\sum_{k\ge0}\left(4(i+1)a^{(m+1)}_{i+1,j}(k)+6(j+1)a^{(m+1)}_{i,j+1}(k)\right)
\Gamma(n-\tfrac{25}{12}-k)\left(\tfrac{21265}{4608}\psi(n-\tfrac{25}{12}-k)+d_1\right)\nonumber\\
{}+\tfrac14S_3\sum_{k\ge0}\left(4(s+1)a^{(m-1)}_{i-1,j}(k)+6(j+1)a^{(m-1)}_{i-2,j+1}(k)\right)
(-1)^{n-k}\Gamma(n+\tfrac{25}{12}-k)\nonumber\\
{}-2(s-2i-1)S_3\sum_{k\ge0}a^{(m-1)}_{i,j}(k)(-1)^{n-k}\Gamma(n-\tfrac{35}{12}-k)
\left(\tfrac{21265}{4608}\psi(n-\tfrac{35}{12}-k)+f_1\right)\nonumber\\
{}-S_3\sum_{k\ge0}\left(8(i+1)a^{(m-1)}_{i+1,j}(k)+6(j+1)a^{(m-1)}_{i,j+1}(k)\right)
(-1)^{n-k}\Gamma(n-\tfrac{95}{12}-k)Q(n-\tfrac{95}{12}-k)\nonumber\\
{}-(f_1-c_1)S_3\sum_{k\ge0}\left(2(i+1)a^{(m-1)}_{i+1,j-1}(k)+6(i+j)a^{(m-1)}_{i,j}(k)\right)(-1)^{n-k}\Gamma(n-\tfrac{35}{12}-k),
\end{gather}
with $s=m-2i-3j$ and $Q(z)=\left(\tfrac{21265}{4608}\right)^2\left(\psi^2(z)+\psi^\prime(z)\right)
+2c_1\left(\tfrac{21265}{4608}\right)\psi(z)+c_2.$

\section{Comments and conclusions}
\begin{enumerate}
\item The conjecture exhibits 17 resurgent terms, all of which have been intensively tested at high precision,
for all actions $m\le8$.
\item The 6 Stokes constants have been determined to better than 1000 digits.
\item Excellent freeware, from  {\tt Pari-GP} in Bordeaux, was vital to this enterprise.
\item First and second derivatives of $\Gamma$  and suppressions by $1/n^{5/6}$ make Richardson acceleration infeasible.
We used systematic matrix inversion.
\item The presence of logarithms in trans-series has been ascribed to resonant actions. 
We find this misleading.  We have shown that a closely analogous second-order problem 
is both resonant and log-free. 
\end{enumerate}
\acknowledgments
We have been guided by helpful advice
from Gerald Dunne and encouraged by the programme and workshops
on {\em  Applicable Resurgent Asymptotics} at the Isaac Newton Institute in Cambridge.

\providecommand{\href}[2]{#2}\begingroup\raggedright\endgroup


\begin{thebibliography}{10}

\bibitem{BK1}
D.~Broadhurst and D.~Kreimer, \emph{{Combinatoric explosion of renormalization
  tamed by Hopf algebra: Thirty loop Pade--Borel resummation}},
  \href{https://doi.org/10.1016/S0370-2693(00)00051-4}{\emph{Phys. Lett. B}
  {\bfseries 475} (2000) 63}
  [\href{https://arxiv.org/abs/hep-th/9912093}{{\ttfamily hep-th/9912093}}].

\bibitem{BK2}
D.~Broadhurst and D.~Kreimer, \emph{{Exact solutions of Dyson--Schwinger
  equations for iterated one loop integrals and propagator coupling duality}},
  \href{https://doi.org/10.1016/S0550-3213(01)00071-2}{\emph{Nucl. Phys. B}
  {\bfseries 600} (2001) 403}
  [\href{https://arxiv.org/abs/hep-th/0012146}{{\ttfamily hep-th/0012146}}].

\bibitem{BDM}
M.~Borinsky, G.V.~Dunne and M.~Meynig, \emph{{Semiclassical trans-series from
  the perturbative Hopf-algebraic Dyson--Schwinger equations: $\phi^3$ QFT in 6
  dimensions}}, \href{https://doi.org/10.3842/SIGMA.2021.087}{\emph{SIGMA}
  {\bfseries 17} (2021) 087}
  [\href{https://arxiv.org/abs/2104.00593}{{\ttfamily 2104.00593}}].

\bibitem{BB}
M.~Borinsky and D.~Broadhurst, \emph{{Resonant resurgent asymptotics from
  quantum field theory}}, \href{https://doi.org/10.1016/j.nuclphysb.2022.115861}{\emph{Nucl. Phys. B} {\bfseries 981} (2022) 115861}
  [\href{https://arxiv.org/abs/2202.01513}{{\ttfamily 2202.01513}}].

\bibitem{polya1937kombinatorische}
G.~P{\'o}lya, \emph{Kombinatorische anzahlbestimmungen f{\"u}r gruppen, graphen
  und chemische verbindungen}, {\emph{Acta mathematica} {\bfseries 68} (1937)
  145}.

\bibitem{Lautrup:1977hs}
B.E.~Lautrup, \emph{On high order estimates in {QED}},
  \href{https://doi.org/10.1016/0370-2693(77)90145-9}{\emph{Phys. Lett. B}
  {\bfseries 69} (1977) 109}.

\bibitem{Connes:1999yr}
A.~Connes and D.~Kreimer, \emph{{Renormalization in quantum field theory and
  the Riemann-Hilbert problem. 1. The Hopf algebra structure of graphs and the
  main theorem}}, \href{https://doi.org/10.1007/s002200050779}{\emph{Commun.
  Math. Phys.} {\bfseries 210} (2000) 249}
  [\href{https://arxiv.org/abs/hep-th/9912092}{{\ttfamily hep-th/9912092}}].

\bibitem{BD}
M.~Borinsky and G.V.~Dunne, \emph{Non-perturbative completion of
  {Hopf}-algebraic {Dyson--Schwinger} equations},
  \href{https://doi.org/10.1016/j.nuclphysb.2020.115096}{\emph{Nucl. Phys. B}
  {\bfseries 957} (2020) 115096}
  [\href{https://arxiv.org/abs/2005.04265}{{\ttfamily 2005.04265}}].

\bibitem{Balduf:2021kag}
P.-H.~Balduf, \emph{{Dyson--Schwinger equations in minimal subtraction}},
  \href{https://arxiv.org/abs/2109.13684}{{\ttfamily 2109.13684}}.

\bibitem{Dyson:1952tj}
F.J.~Dyson, \emph{{Divergence of perturbation theory in quantum
  electrodynamics}}, \href{https://doi.org/10.1103/PhysRev.85.631}{\emph{Phys.
  Rev.} {\bfseries 85} (1952) 631}.

\bibitem{Fisher_1978}
M.E.~Fisher, \emph{{Yang--Lee} edge singularity and ${\ensuremath{\phi}}^{3}$
  field theory}, \href{https://doi.org/10.1103/PhysRevLett.40.1610}{\emph{Phys.
  Rev. Lett.} {\bfseries 40} (1978) 1610}.

\bibitem{ecalle1981fonctions}
J.~{\'E}calle, \emph{Les fonctions r{\'e}surgentes}, vol.~1, Universit{\'e} de
  Paris-Sud, D{\'e}partement de Math{\'e}matique, B{\^a}t. 425 (1981),
\href{http://sites.mathdoc.fr/PMO/PDF/E_ECALLE_81_05.pdf}
{\emph{Les alg{\`e}bres de fonctions r{\'e}surgentes}}.

\end{thebibliography}
\end{document}